\documentclass[usegraphicx,usenatbib, useAMS]{mn2e}
\usepackage{amsfonts}
\usepackage{amsmath}
\usepackage{amssymb}
\usepackage{comment}
\usepackage{subfloat}
\usepackage{subfig}
\usepackage[usenames,dvipsnames]{color}
\usepackage{times}
\def\lesssim{\mathrel{\hbox{\rlap{\hbox{\lower4pt\hbox{$\sim$}}}\hbox{$<$}}}}
\def\gtrsim{\mathrel{\hbox{\rlap{\hbox{\lower4pt\hbox{$\sim$}}}\hbox{$>$}}}}

\title[Major mergers at z$\simeq$2]{Major mergers are not significant drivers of star formation or morphological transformation around the epoch of peak cosmic star formation}

\author[E K. Lofthouse et al.]
{E. K. Lofthouse$^{1}$\thanks{E-mail: e.k.lofthouse@herts.ac.uk}, S. Kaviraj$^{1}$, C.J.Conselice$^{2}$, A. Mortlock$^{2}$ and W. Hartley$^{2}$\\
$^{1}$Centre for Astrophysics Research, University of Hertfordshire, College Lane, Hatfield, Herts AL10 9AB, UK \\
$^{2}$University of Nottingham, School of Physics and Astronomy, Nottingham NG7 2RD}

\begin{document}

\date{\today}

\pagerange{\pageref{firstpage}--\pageref{lastpage}}

\maketitle

\label{firstpage}

\begin{abstract}

We investigate the contribution of major mergers (mass ratios $>1:5$) to stellar mass growth and morphological transformations around the epoch of peak cosmic star formation ($z\sim2$). We visually classify a complete sample of massive (M $>$ 10$^{10}$M$_{\odot}$) galaxies at this epoch, drawn from the CANDELS survey, into late-type galaxies, major mergers, spheroids and disturbed spheroids which show morphological disturbances. Given recent simulation work, which indicates that recent ($<$0.3-0.4 Gyr) major-merger remnants exhibit clear tidal features in such images, we use the fraction of disturbed spheroids to probe the role of major mergers in driving morphological transformations. The percentage of blue spheroids (i.e. with ongoing star formation) that show morphological disturbances is only 21 $\pm$ 4 per cent, indicating that major mergers are not the dominant mechanism for spheroid creation at $z\sim2$ - other processes, such as minor mergers or cold accretion are likely to be the main drivers of this process. We also use the rest-frame U-band luminosity as a proxy for star formation to show that only a small fraction of the star formation budget ($\sim$3 per cent) is triggered by major mergers. Taken together, our results show that major mergers are not significant drivers of galaxy evolution at $z\sim2$. 
\end{abstract}


\section{Introduction}

  The processes which drive star formation and the morphological transformation of galaxies at high redshift ($z>1$) remain the subject of much debate. Since the cosmic star formation rate (SFR) peaks at z $\simeq$ 2 \citep{Madau98}, this is the epoch at which a significant proportion of the stellar mass in today's  galaxies formed. This is also the epoch at which the morphological mix of the Universe changes rapidly, settling into something that resembles the local Hubble sequence \citep{Lee13,Mortlock13}. However, the principal mechanisms responsible for this evolution remain unclear.

  Major mergers have traditionally been thought to be significant drivers of galaxy evolution, triggering strong star formation, AGN, the growth of black holes and inducing the morphological transformations which create spheroidal galaxies. Simulations have shown that mergers between two spirals create spheroidal systems \citep{Negroponte83, DeLucia06, Springel05}, supporting the hypothesis that the spheroid formation may be largely driven by such mergers \citep[e.g.][]{AvilaReese14}. Major mergers are also clearly capable of inducing strong starbursts \citep{Joseph84,Hernquist89, Mihos96}, suggesting that a large fraction of the stellar mass growth may be due to this process.

  However, recent work indicates that major mergers may not be as important as other processes, such as minor mergers or cold accretion, in driving galaxy evolution at early epochs, e.g. \citet{Dekel09,Genzel08,Kaviraj11, Conselice13, Dullo13,Graham15}. The contribution of major mergers to the cosmic star formation budget at $z \sim2$ appears minimal \citep{Kaviraj13c}. In a recent study of a small sample of massive galaxies, \citet{Kaviraj13a} suggest that $\sim$50 per cent of blue spheroids at $z \sim 2$ do not show tidal features indicative of major mergers, and those that do only show modest increases in their specific star formation rates (sSFRs).

  \citet{Williams14} have similarly argued that compact star forming galaxies at z $\sim $ 3 are likely to have evolved into compact early-type galaxies at z $\sim $ 1.6 via accretion processes and not by major mergers. This was determined from the lack of extended haloes in the light distribution. The S\'ersic light profiles of these objects prefer one-component best fits rather than the two-component fits expected for merger remnants, in which a central starburst is expected to be superimposed on an underlying stellar distribution \citep{Wuyts10}. 

  In this paper, we investigate the observational evidence for major mergers in star-forming spheroidal galaxies and their contribution to the total cosmic star formation at z$\sim$2 using a sample over seven times larger than in any previous works. We also extend this work by studying the variation in fraction of disturbed blue spheroids as a function of redshift. 

  This paper is structured as follows. In Section~\ref{data}, we describe the HST data employed in this study. Section~\ref{class} discusses the visual morphological classification of our galaxy sample and the identification of major-merger remnants. In Section~\ref{frac}, we explore the role of major mergers in triggering the creation of early spheroids, while in Section~\ref{budget} we study the contribution of this process to the star formation budget around the epoch of peak star formation. We summarize our findings in Section~\ref{summary}.

\section{Data}\label{data}

The Cosmic Assembly Near-infrared Deep Extragalactic Legacy Survey (CANDELS; \citealt{Grogin11}, \citealt{Koekemoer11}) is a Multi Cycle Treasury Programme using the Wide Field Camera 3 (WFC3) and the Advanced Camera for Surveys (ACS) on the Hubble Space Telescope (HST). CANDELS covers a total of 800 acrmin$^2$ divided into five different fields: GOODS-N, GOODS-S, EGS, COSMOS and UDS. This is comprised of CANDELS/Deep which images GOODS-N and GOOD-S to a 5$\sigma$ point source depth of $H_{AB}$ = 27.7 and CANDELS/Wide which covers all field to a 5$\sigma$ depth of $H_{AB}$ = 26.3 in a 1 arcsec aperture.

In this paper we use the data from CANDELS UDS (C-UDS) and CANDELS GOODS-S. All magnitudes are quoted this paper are AB magnitudes. The area covered by C-UDS and GOODS-S are approximately 0.06 and 0.05 deg$^2$ respectively. In addition to the U-band CFHT data, B, V, R, i, z-band SXDS data and J, H and K-band data from UKDISS available for the whole UDS field, C-UDS also includes F606W and F814W data from the ACS, H$_{160}$ and J$_{125}$-band HST WFC3 data from CANDELS, Y and Ks bands taken as part of HUGS \citep{Fontana14} and 3.6 and 4.5 $\mu$m from SEDS \citep{Ashby13}. For the CANDELS GOODS-S region photometry consists of: two U bands (one from MOSIAC II image on cerro Tololo Inter-American Observatory (CTIO) Blanco telescope and one with VIsible MultiObject Spectrograph (VIMOS) on the VLT) ACS F435W, F606W, F775W, F814W, F850LP data, WFC3 F098W, F105W, J$_{125}$ and H$_{160}$-band data, two sets of K-band photometry (one from ISAAC instrument on the VLT and one from the HUGS programme using HAWK-I on the VLT) and all four IRAC bands from SEDS.

Stellar masses are derived from a spectral energy distribution (SED) fitting technique, using a combined best-fit and Bayesian approach. Model SEDs are constructed from the stellar population models of \citet{Bruzual03} and using exponentially declining star formation histories with various ages, metallicities and dust extinctions. These SEDs are fitted to the photometric data under the assumption of a Chabrier IMF \citep{Chabrier03}. 
For further details see \citet{Mortlock15} and \citet{Ownsworth14}. The photometric redshifts are produced using EAZY \citep{Brammer08} by fitting template spectra to the optical and near infra-red bands. For a full description see \citet{Hartley13, Mortlock13}. Since we are interested in the intrinsic U-band luminosities, the U band magnitudes that are used for the star-formation budget analysis in Section~\ref{budget} have been corrected for internal dust extinction using E(B-V) values. These values are derived from the SED fitting which uses the reddening law of \citet{Calzetti00} and provides the best fit value for the internal dust in each galaxy. We use a total-to-selective extinction ratio of R$_{U}$= 4.904 derived from the values given in \citet{Schlegel98}.
For more details on the dust correction, see \citet{Mortlock15}.

\section{Morphological Classification}\label{class}

\begin{figure}
  \centering
  \includegraphics[trim=0cm 3cm 0cm 0cm, clip, width=0.45\textwidth]{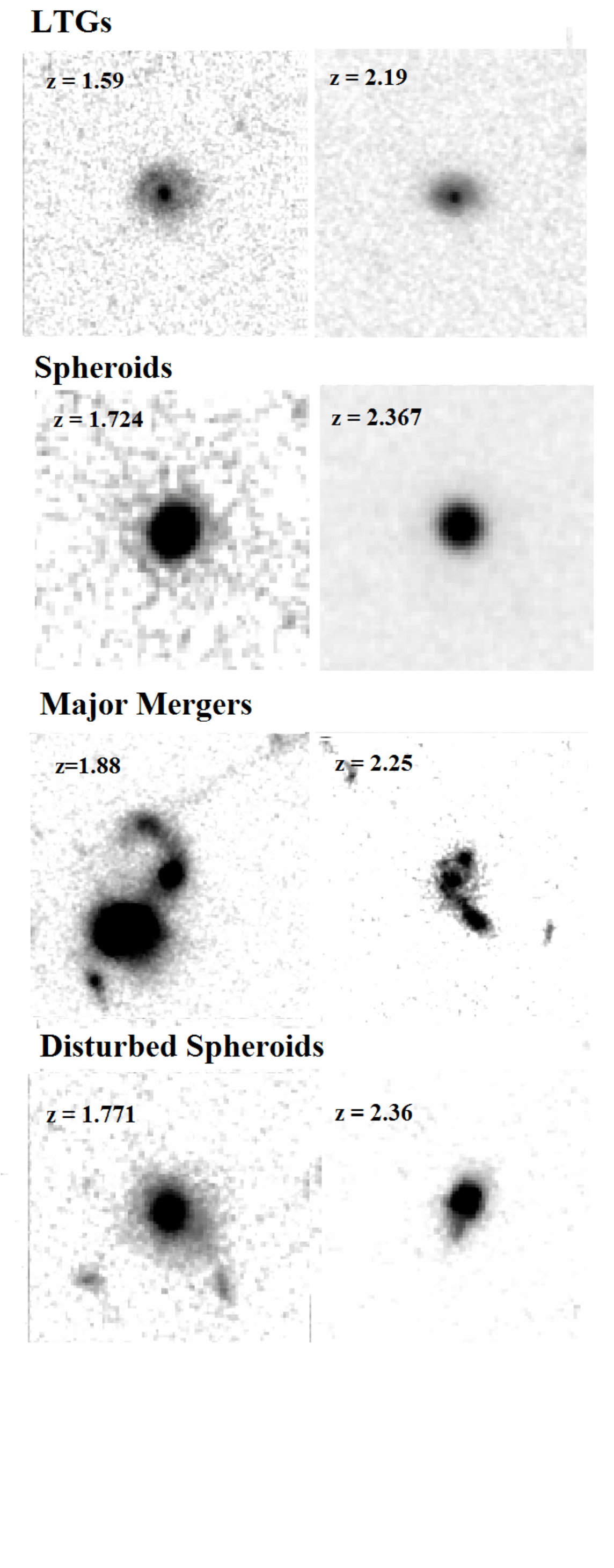}
  \caption{Example H-band images, taken by HST WFC3 camera, of the different morphological classes in our sample, in both the low and high redshift bins. The images are 2.5'' across. }
  
  \label{images}
  \end{figure}

  We construct a sample of 595 galaxies with m$_{H160}<23$, M$_{*}$ $>$ $ 10 ^{10} M_{\odot} $, and redshifts in the range 1.5 $<$ z $<$ 2.5. This sample is mass-complete at these magnitudes. Images taken by WFC3 in the near infrared ($H$-band) are used, which trace the restframe optical properties at  z$\sim$2. This allows the study of the overall morphological structure of the galaxy. We visually classify the $H$-band images into non-interacting spheroids (154 galaxies), non-interacting late-type galaxies (LTGs, 287 galaxies), mergers (117 galaxies) and disturbed spheroids (37 galaxies). Major mergers are defined as a galaxy with two distinct interacting cores and tidal bridges between them. Disturbed spheroids contain a single core but show tidal features indicating a recent major merger has occurred (37 galaxies).  Figure \ref{images} shows examples of typical galaxies for each category.

  These visual classifications agree well with the morphological fractions in other works. For example, using the classifications from \citet{Kartaltepe15}, where \textgreater{}2/3 classifiers agree, yields an LTG fraction of 42 per cent and a spheroid fraction of 29 per cent  \citep{Peth16}.

  \citet{Kaviraj13a} have recently used hydrodynamical simulations to model the surface brightness of tidal features in CANDELS-like images. They simulated mergers of galaxies with various mass ratios and showed that a `major' merger (mass ratio of galaxies greater than 1:5) occurring within the previous $\sim$0.5 Gyr would produce distinct tidal features that are definitely detectable at the depth of the CANDELS images up to z$\sim$2.5. Tidal features from galaxy mergers with mass ratio less than 1:5 will not be detectable at the image depths and redshifts used in this work. Thus all systems in our `disturbed spheroids' class are likely to have experienced a recent major merger.

\section{Do major mergers dominate spheroid formation?}\label{frac}

  We begin by exploring the role of major mergers in creating spheroids in the early Universe, using a method that closely follows \citet{Kaviraj13a}. For this exercise, we focus on \emph{blue} spheroids (based on the dust-corrected rest-frame colours). Galaxies that are currently forming stars are blue due to the presence of hot, short-lived stars \citep{Searle73}. Hence, morphologically disturbed blue spheroids in our sample are likely to be forming a significant fraction of their stellar mass due to a recent major merger. Note that the timescale ($\sim$ 0.5-1 Gyr) over which tidal features fade is comparable to the median timescales over which galaxies redden \citep{Kaviraj13a, Lotz08}. Therefore, it is unlikely that any spheroids formed by major mergers could have undetectable tidal features and still be blue.

  Studying blue spheroids at this particular epoch offers key insights into the creation of the spheroidal population for two reasons. Firstly, the morphological mix of the Universe changes dramatically at this epoch, with the spheroid population being built over a short space of cosmic time \citep{Buitrago13, Conselice14b}. Hence, blue spheroids at this epoch are being observed in the act of formation. Secondly, spheroids are known to form their stars quickly given their high stellar [$\alpha$/Fe] ratios, with a star formation timescale shorter than roughly 1 Gyr \citep[e.g.][]{Trager00,Thomas05}. Therefore, when we look at blue spheroids at this epoch, the star formation we are seeing is likely to be the principal star formation episode which creates a significant fraction of the stellar mass in these galaxies (since multiple episodes would result in the star formation history becoming too extended, diluting the stellar [$\alpha$/Fe] ratios to lower than has been observed). Hence, this is the epoch at which the morphological mix of the Universe is changing. 

\begin{figure}  \includegraphics[width=0.5\textwidth]{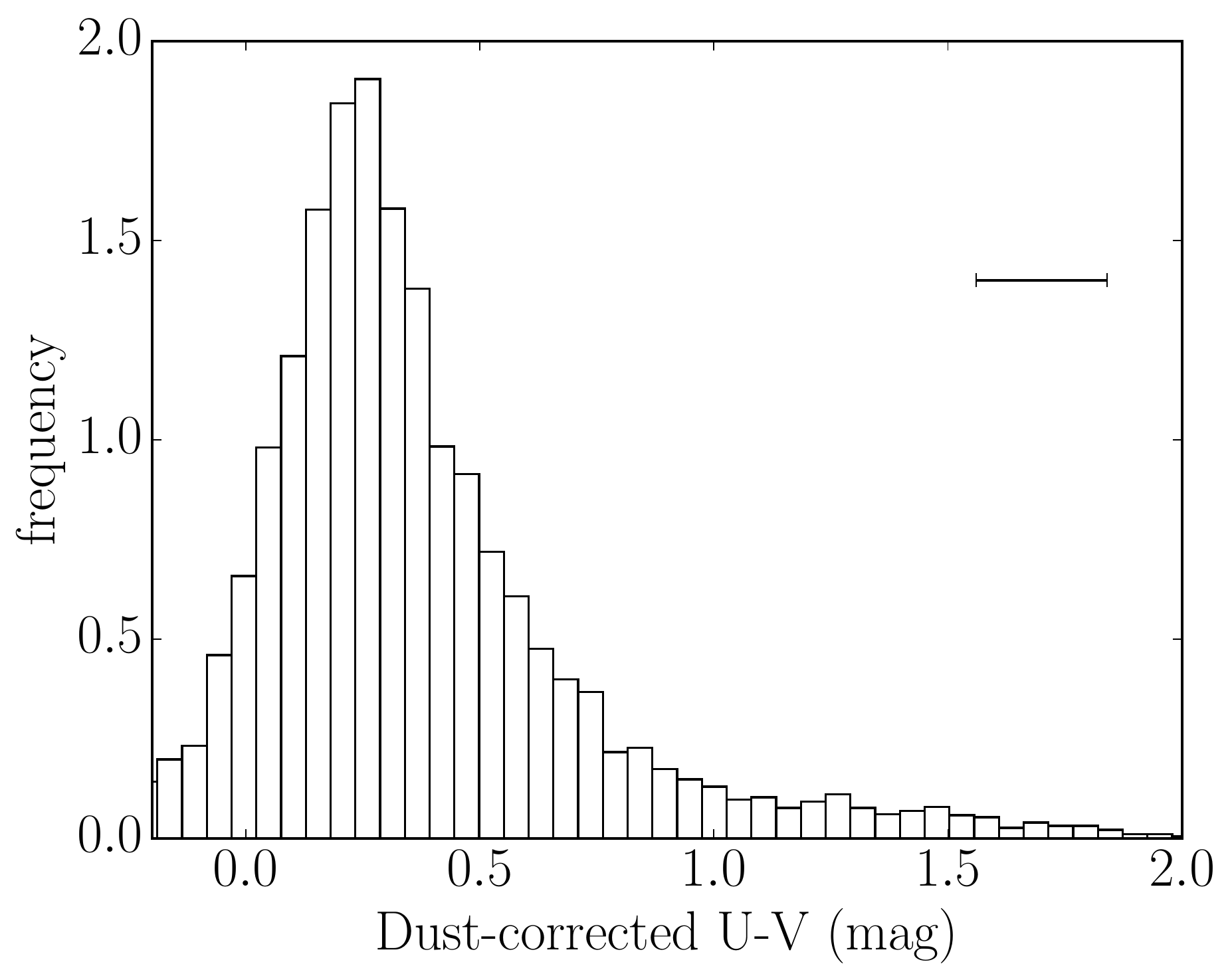}\\
  \caption{The normalised distribution of dust-corrected U-V colours for galaxies in GOODS and UDS with redshifts 1.5 \textless{} z \textless{} 2.5. The large peak on the left shows the blue galaxies, with a small tail of red galaxies on the right. U-V = 1.2 is used as a cut between red and blue galaxies to select blue spheroids. The median error on the U-V colours is shown in the top right of the plot. See text for details.}\label{ColourCut}
  \end{figure}

  Blue galaxies are selected using rest-frame dust-corrected U-V colours (Figure~\ref{ColourCut}), which shows the expected large peak of blue galaxies and an extended distribution towards red galaxies. We define the cut-off between `red' and `blue' galaxies to be U-V = 1.2. Using this blue spheroid sample we find that only 21 $\pm$ 4 per cent of these systems show morphological disturbances. Our results show that only a small fraction of spheroids that are forming at this epoch are likely being created via the major-merger process. There is a possibility that some spheroids are formed more slowly over time by the addition of mass in their outer regions (e.g by minor merging) rather a larger, galaxy wide burst of star formation as studied here. However, if this is true it would result in an even smaller fraction of spheroids being formed via major mergers than the 21 per cent found in this work, reinforcing our conclusion that major mergers are not the dominant mechanism for spheroid creation.  

\begin{figure}
  \includegraphics[width=0.5\textwidth]{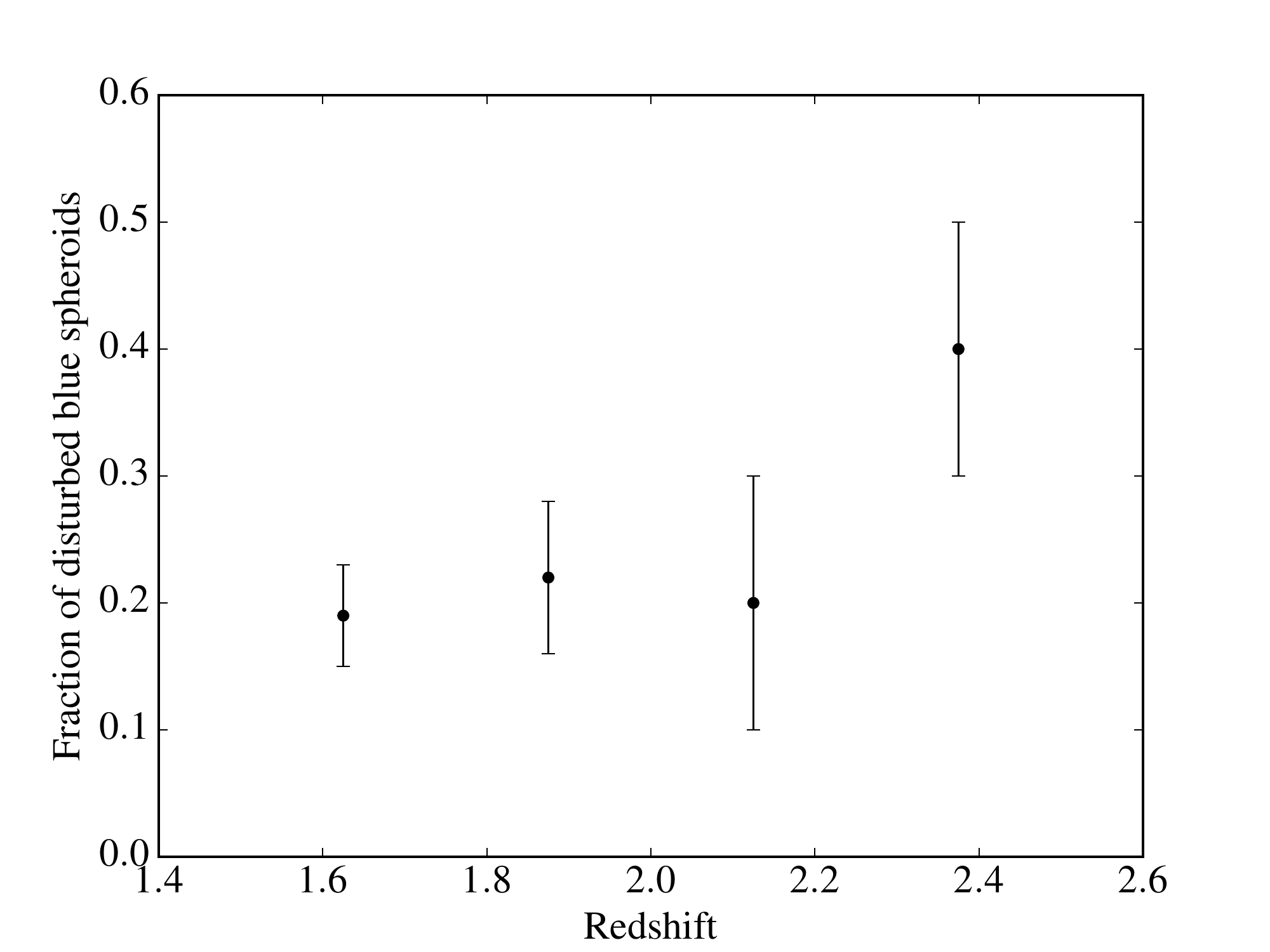}\
  \caption{Fraction of blue spheroids with morphological disturbances binned by redshift in ranges 1.5 \textless{} z \textless{} 1.75, 1.75 \textless{} z \textless{} 2, 2 \textless{} z \textless{} 2.25, 2.25 \textless{} z \textless{} 2.5. There is a possible increase in the fraction of blue spheroids with morphological disturbances as redshift increases but the error bars are large enough that the trend is also consistent with a flat merger rate evolution. The points are plotted at the mid-point of each bin with errors that are calculated using standard error propagation equations.
  }\label{zComb}
  \end{figure}

  We proceed by splitting our sample into four redshift bins and calculate the percentage of blue spheroids with morphological disturbances in each of these redshift ranges. The results are shown in Figure~\ref{zComb}. This indicates a possible increase in the fraction of blue spheroids with morphological disturbances to higher redshift. While at lower redshifts, 1.5 \textless{} z \textless{} 1.75 only 19 $\pm$ 4 per cent of blue spheroids have morphological disturbances whereas at the highest redshifts, 2.25 \textless{} z \textless{} 2.5, the fraction increases to 40$\pm$10 per cent. While this may indicate an increase in the merger rate with increasing redshift \citep[see e.g.][]{Wolf05,Bluck12}, the errors are large and the result is also consistent with a flat merger rate in our redshift range of interest. \citep[see e.g.][]{Kaviraj15}. Further work using larger samples of galaxies at the higher end of our redshift range is desirable.

\section{Do major mergers contribute significant additional mass growth by enhancing star formation?}\label{budget}

  We study the star formation activity in our galaxy population and the role of ongoing major mergers in driving stellar mass growth using rest-frame U-band luminosities derived from the SED fitting described above. The U-band luminosity is dominated by emission from young massive stars and hence can be effectively used to trace recent star formation  \citep{Hopkins03}. It has the advantage of suffering from less dust attenuation than the more commonly used far-UV SFR indicators \citep{Moustakas06}. However, there is still some dust attenuation so we correct our U-band luminosities using dust estimates obtained from the best-fit SEDs. 

\begin{figure}
  \centering
  \begin{tabular}{@{}lc@{}}
  \hline
      Morphology     &    Fraction of U-band Luminosity Budget \\%
  \hline
      Non-interacting LTGs    & 0.53 $\pm$ 0.10   \\
      Non-interacting spheroids    & 0.19 $\pm$ 0.04   \\
      Mergers     & 0.22 $\pm$ 0.04   \\
      Disturbed spheroids  & 0.058 $\pm$ 0.014 \\
  \hline
  \end{tabular}
  \includegraphics[trim = 0cm 3cm 0cm 2cm, clip, width=0.45\textwidth]{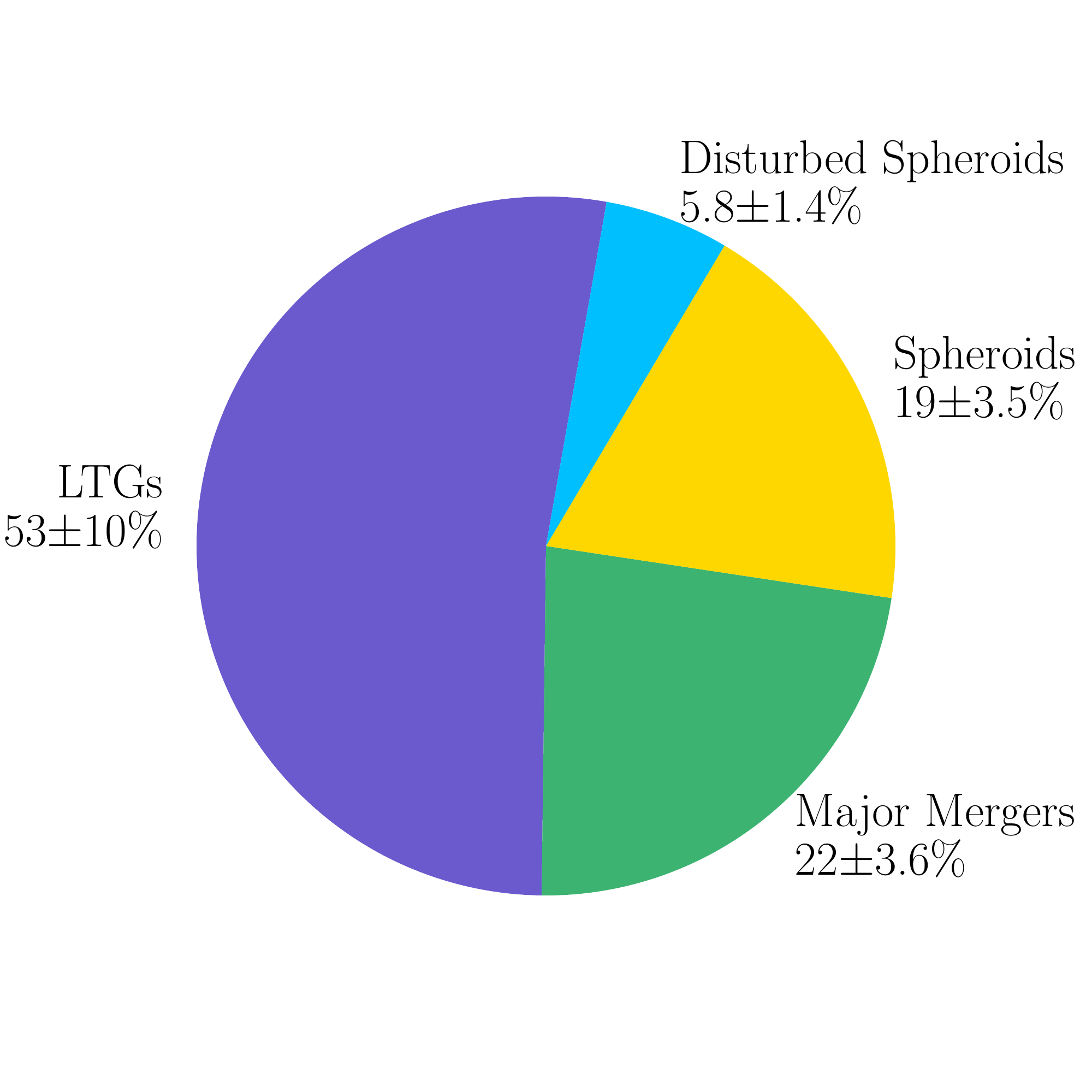}
  \caption{Top: The fraction of U-band Luminosity from massive galaxies ($M_{*} $\textgreater$ 10 ^{10} M_{\odot} $ at z$\sim2$) split by morphological type, LTG = Late-type galaxies, spheroids, disturbed spheroids and major mergers. Bottom: Pie chart representation of these fractions showing the distribution of the overall U-band luminosity budget. Half of all the luminosity is in LTGs, 53 $\pm$ 8 per cent with only 29 $\pm$ 3 per cent in galaxies associated with a merger. The errors are calculated using the standard error propagation formulas to combine the uncertainties on the dust-corrected U-band luminosity for each galaxies along with the Poisson errors in the number counts.} 
  \label{uBudget}
  \end{figure}

  We calculate the total U-band luminosities in each morphological type: LTGs, spheroids, major mergers and disturbed spheroids. The percentage of total luminosity for each morphological type is shown in Figure~\ref{uBudget}. The errors are calculated from the standard Poisson errors on the number counts in each morphology combined with the uncertainty on the dust-corrected U-band luminosity. We find 72 per cent of the luminosity is emitted by non-interacting galaxies. The majority of this emission is from LTGs, with 53 $\pm$ 10 per cent of the total budget and the rest is from the spheroids with 19 $\pm$ 4 per cent. The remaining $\sim$28 per cent of the total budget is from galaxies which are currently undergoing a major merger (22 $\pm$ 2 per cent) or are disturbed spheroids due to a recent major merger (5.8 $\pm$ 1.4 per cent).  These results are in agreement with previously published percentages for the distribution of star formation in different morphologies. For example, at the same redshifts of z $\sim$ 2, \citet{Kaviraj13c} found that 55 per cent of the cosmic star formation rate density (SFRD) is in non-interacting LTGs and only 27 per cent is from major mergers. 

\begin{figure}
  \centering
  \includegraphics[width=0.5\textwidth]{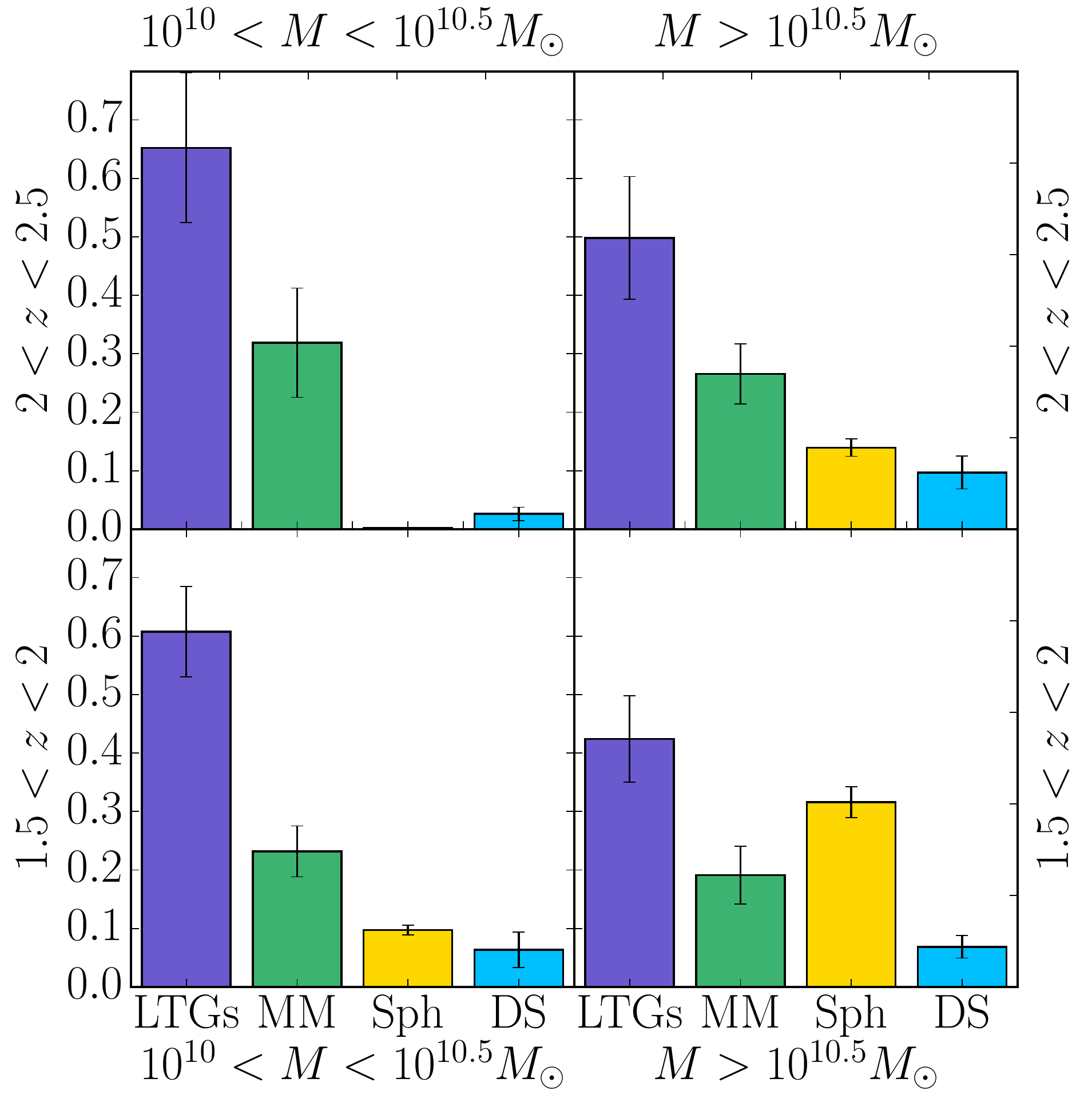}
  \caption{The U-band Luminosity budget in terms of galaxy morphology (LTGs = late-type galaxies; MM = Major Mergers, Sph = Spheroids, DS = Disturbed Spheroids) split into two mass bins and two redshift bins. At higher redshifts, the percentage of star formation hosted by major mergers increases while the percentage hosted by spheroids decreases. On the other hand, moving to higher masses shows a larger percentage from the spheroid population. }
  \label{Upie2by2}
  \end{figure}

  To investigate whether the U-band luminosity budget shows any variation by mass or redshift, we split the sample into four bins combining low and high redshifts (1.5 \textless{} z \textless{} 2 and 2 \textless{} z \textless{} 2.5) and low and high masses ($ 10 ^{10} M_{\odot} $ \textless{} M$_{*}$ \textless{} $ 10 ^{10.5} M_{\odot} $ and M$_{*}$  \textgreater{} $10 ^{10.5} M_{\odot}$). The fraction of the luminosity for each morphological type is calculated and shown in Figure~\ref{Upie2by2}. In the higher redshift range, there is a larger percentage of luminosity in major mergers and a lower percentage in spheroids compared to the low redshift range. At higher masses the percentage of luminosity from LTGs is lower than at lower masses while the percentage in both non-interacting spheroids and disturbed spheroids is higher. This is likely due to the larger number fraction of spheroids at high masses. 39 $\pm$ 5 per cent of galaxies in the high mass bin are classified as spheroids compared to 11 $\pm$ 2 per cent in the lower mass bin.

\subsection{Enhancement due to major mergers}\label{enhancement}

  The previous results show that major mergers are not the main source of the U-band luminosity. However, we note that the ratio of average specific SFR for all major mergers (ongoing mergers and disturbed spheroids) to LTGs is $\sim$1.1:1. This shows that there is little enhancement in the U-band luminosity when a major merger occurs and hence there is no significant increase in the SFR due to a major merger. This is supported by recent results from simulations showing little star formation enhancement due to high-redshift mergers \citep{Fensch16}.

   The fraction of the U-band luminosity budget that is contained in merging systems is an upper limit to the U-band luminosity that is attributable to the major merger process. This is because even if the merger were not taking place the progenitors would still be forming stars via other means as seen by the non-zero SFRs in the non-interacting galaxies. Indeed, given that the ratio of U-band luminosities between mergers and LTGs is so similar implies that the star formation actually triggered by major merging is small.

   Combining the enhancement by major mergers determined from this ratio (1.1 - 1) with the result in Figure~\ref{uBudget} that 28 per cent of the luminosity is in mergers suggests that only 3 per cent ($ 28 $ per cent$ * 0.1/1.1$) of the cosmic U-band luminosity at z $\sim$ 2 is directly driven by the major merger process. It is worth noting that \citet{Kaviraj13c} performed the same calculation on a much smaller number of galaxies and narrower mass range but arrived at a similar result. They found a ratio of $\sim$2.2:1 for the sSFR between mergers and LTGs and concluded that $<$15 per cent of the cosmic SFRD is due to the major merger process. These conclusions are supported by previous observational work which indicates that a high fraction of star-forming systems at this epoch are not major mergers but instead are driven by other processes such as secular evolution or minor mergers \citep[e.g.][]{Genzel08,Tacconi10}.

\section{Summary}\label{summary}

  We have studied a sample of 595 massive (M$_{*}$ \textgreater{} $10^{10} M_{\odot} $) galaxies in order to investigate the role of major mergers in creating spheroids and driving star formation around the epoch of peak cosmic star formation. Noting that simulations indicate that a recent major merger will leave clear morphological disturbances at the surface brightness limit of our images, we have studied the incidence of tidal features around blue spheroids (i.e. spheroids that are currently star forming) around the epoch of peak cosmic star formation. Given that this is the epoch around which the bulk of the spheroid population is created, this enables us to explore the role of major mergers in driving the production of these galaxies. Our results indicate that only 21 $\pm$ 4 per cent of the blue spheroids show morphological disturbances. The blue disturbed fractions show an increase with redshift (suggesting an increase in the merger rate with redshift), although the error bars are large enough that the trend is also consistent with a flat merger rate evolution in our redshift range of interest. Our analysis indicates that major mergers are not the principal driver of spheroid formation around the epoch of peak star formation. Therefore other mechanisms, such as minor mergers (which may drive disc instabilities) could be the dominant mechanism that triggers morphological transformation in the early Universe \citep{Kaviraj14}. 

  In the second half of our study we have used the rest-frame U-band luminosity to probe the role of major mergers in driving star formation in the early Universe. We have shown that only a small fraction of the U-band luminosity budget is from galaxies involved in a major merger, $\sim$30 per cent. By splitting the sample into 4 bins by mass and redshift we found that the luminosity budget shows a higher percentage of major mergers at higher redshifts. The contribution from spheroids decreases with redshift and increases with stellar mass. We have argued that the fraction of the U-band luminosity budget hosted by major mergers is an upper limit, as the merging progenitors will likely still contain ongoing star formation even if they were not merging. The star formation activity that is directly attributable to the major merger process is therefore the star formation enhancement observed in merging systems. We have compared the enhancement due to mergers by studying the the ratio of the average specific SFR in mergers to that in LTGs, and find this to be $\sim$1.1:1. Combining this with the results for the U-band luminosity  budget indicates that only 3 per cent of the cosmic star formation budget at z $\sim$ 2 is due to the major merger process. This is lower than the percentage of cosmic SFRD in mergers found by \citet{Kaviraj13c} who used a similar method albeit with a much smaller sample of galaxies, but the results both indicate the small contribution of major mergers to the SFRD at these epochs and are supported by theoretical work \citep[e.g.][]{Dekel09,Keres09} and observations \citep[e.g.][]{Williams14}.
 
  Overall, our results show that major-merging is only a minor player in galaxy evolution around the epoch of peak cosmic star formation.

\section*{Acknowledgements}

We are grateful to the anonymous referee for many constructive comments that improved the quality of the original manuscript. SK is supported by STFC grant ST/M001008/1 and acknowledges a Senior Research Fellowship from Worcester College Oxford. EKL acknowledges support from the UK’s Science and Technology Facilities Council [grant number St/K502029/1]. AM acknowledges funding from the STFC and a European Research Council Consolidator Grant (P.I. R.McLure).

\bibliographystyle{mn2e}
\bibliography{BibTex}
\bsp

\label{lastpage}

\end{document}